\documentclass[aps,prl,10pt,superscriptaddress,notitlepage,twocolumn,amsmath,amssymb]{revtex4-2}

\usepackage{bm}
\usepackage{braket}

\usepackage{graphicx}
\graphicspath{{./figures/}}

\DeclareMathOperator{\tr}{tr}

\begin{document}

\title{Closed-loop optimization of fast trapped-ion shuttling with sub-quanta excitation}

\author{Jonathan D. Sterk}
\affiliation{Sandia National Laboratories, Albuquerque NM, USA 87185}
\email{jdsterk@sandia.gov}
\author{Henry Coakley}
\affiliation{Sandia National Laboratories, Albuquerque NM, USA 87185}
\author{Joshua Goldberg}
\affiliation{Sandia National Laboratories, Albuquerque NM, USA 87185}
\author{Vincent Hietala}
\affiliation{Sandia National Laboratories, Albuquerque NM, USA 87185}
\author{Jason Lechtenberg}
\affiliation{Sandia National Laboratories, Albuquerque NM, USA 87185}
\author{Hayden McGuinness}
\affiliation{Sandia National Laboratories, Albuquerque NM, USA 87185}
\author{Daniel McMurtrey}
\affiliation{Sandia National Laboratories, Albuquerque NM, USA 87185}
\author{L. Paul Parazzoli}
\affiliation{Sandia National Laboratories, Albuquerque NM, USA 87185}
\author{Jay Van~Der~Wall}
\affiliation{Sandia National Laboratories, Albuquerque NM, USA 87185}
\author{Daniel Stick}
\affiliation{Sandia National Laboratories, Albuquerque NM, USA 87185}

\begin{abstract}
    Shuttling ions at high speed and with low motional excitation is essential for realizing fast and high-fidelity algorithms in many trapped-ion based quantum computing architectures.
    Achieving such performance is challenging due to the sensitivity of an ion to electric fields and the unknown and imperfect environmental and control variables that create them.
    Here we implement a closed-loop optimization of the voltage waveforms that control the trajectory and axial frequency of an ion during transport in order to minimize the final motional excitation.
    The resulting waveforms realize fast round-trip transport of a trapped ion across multiple electrodes at speeds of $0.5$~electrodes/$\mu$s ($35~\text{m/s}$) with a maximum of $0.36\pm0.08$ quanta gain.
    This sub-quanta gain is independent of the phase of the secular motion at the distal location, obviating the need for an electric field impulse or time delay to eliminate the coherent motion.
\end{abstract}

\maketitle

\section{Introduction}
\label{sec:intro}
Trapped ions are a leading technology platform for quantum computing due to their long coherence times and high-fidelity quantum operations. 
While current trapped-ion based quantum computers and simulators employ tens of trapped ions~\cite{wrightBenchmarking2019,bermudezAssessing2017}, practical quantum computation may ultimately require upwards of $10^6$ ions \cite{lekitschBlueprint2017}. 
The earliest proposed architecture for scaling trapped ion systems relies on ion transport for connecting qubits and is known as the Quantum Charge Coupled Device (QCCD) architecture~\cite{kielpinskiArchitecture2002}.
All transport primitives required for moving ions within the QCCD architecture (\emph{i.e.}, splitting, shuttling, merging and reordering) have been demonstrated in small systems~\cite{schulzOptimization2006, homeComplete2009, blakestadNeargroundstate2011, rusterExperimental2014,kaufmannDynamics2014, pinoDemonstration2021,hilderFaulttolerant2021,kaushalShuttlingbased2020}.

A time-budget analysis of experiments on the QCCD architecture illustrates that ion-shuttling can consume a significant fraction of the total algorithm operation time~\cite{bowlerArbitrary2013,pinoDemonstration2021,kaushalShuttlingbased2020}, thus highlighting the need for fast transport.
Shuttling must also not substantially excite ion motion, since the motional modes mediate spin--spin interactions for entangling gates and coherent excitation on the order of single quanta can lead to a loss in fidelity~\cite{ruzicEntanglinggate2021a}.
For a many-ion array, these requirements must be achieved in spite of imperfections in the environment and control system. 

An important metric for a shuttling-based architecture is the number of electrode-lengths shuttled per second, as this represents the time to transport an ion to an independent trapping site. 
Earlier efforts~\cite{bowlerCoherent2012,waltherControlling2012} have demonstrated comparable or faster transport using high speed voltage waveform generators, albeit at lower electrode/s rates due to the larger electrode sizes.
These approaches require precise timing in order to realize shuttling with low excitation, such as synchronizing the transport with the axial frequency or through the use of a diabatic electric field impulse at the correct phase of axial motion to remove the excitation.

Theoretical research in shuttling protocols have used optimal control theory~\cite{schulzOptimization2006,furstControlling2014} and invariant-based engineering to realize shortcuts to adiabaticity~\cite{torronteguiFast2011, chenLewisRiesenfeld2011,chenOptimal2011,guery-odelinShortcuts2019}. 
Such protocols can in principle yield shuttling solutions that transport an ion with no motional excitation with transport times much shorter than a period of the axial motion.
These techniques can be utilized to generate large coherent states of motion~\cite{alonsoGeneration2016a} as well as be extended for multi-ion chains~\cite{palmeroFast2014a}, anharmonic traps~\cite{zhangOptimal2016}, and time-dependent axial frequencies~\cite{tobalinaFast2017}.
While these theoretical results are promising, these protocols rely upon accurate physical modelling as well as accurate realization of the controls. 

Instead of relying upon the accuracy of our models and their realization, we perform a closed-loop optimization where the voltage waveform is optimized against experimental runs.
In this manner, a highly accurate model of the experimental apparatus is unnecessary as it is the experimental performance of the voltage waveform that is being optimized.
The resulting waveform is able to account for imperfections that excite ion motion during transport, such as fabrication and geometry differences across a device, imperfect simulations, background electric fields, and disparities in filter components that modify the temporal properties of the voltage waveforms.
The technique we describe constitutes a tool that could be used to tune shuttling protocols for many ions within a trapping array and is agnostic to deviations from the model resulting from environmental effects and variations in the fabrication process.
Furthermore we note that this procedure could be appliciable to tuning up other experimental quantum technologies, such as neutral-atom quantum computation~\cite{stuartSingleatom2018} and atom interferometry~\cite{steffenDigital2012,corgierFast2018,amriOptimal2019,duspayevTractor2021}.

\section{Results}
\subsection{Optimization procedure}
\begin{figure*}
    \centering
    \includegraphics[width=\textwidth]{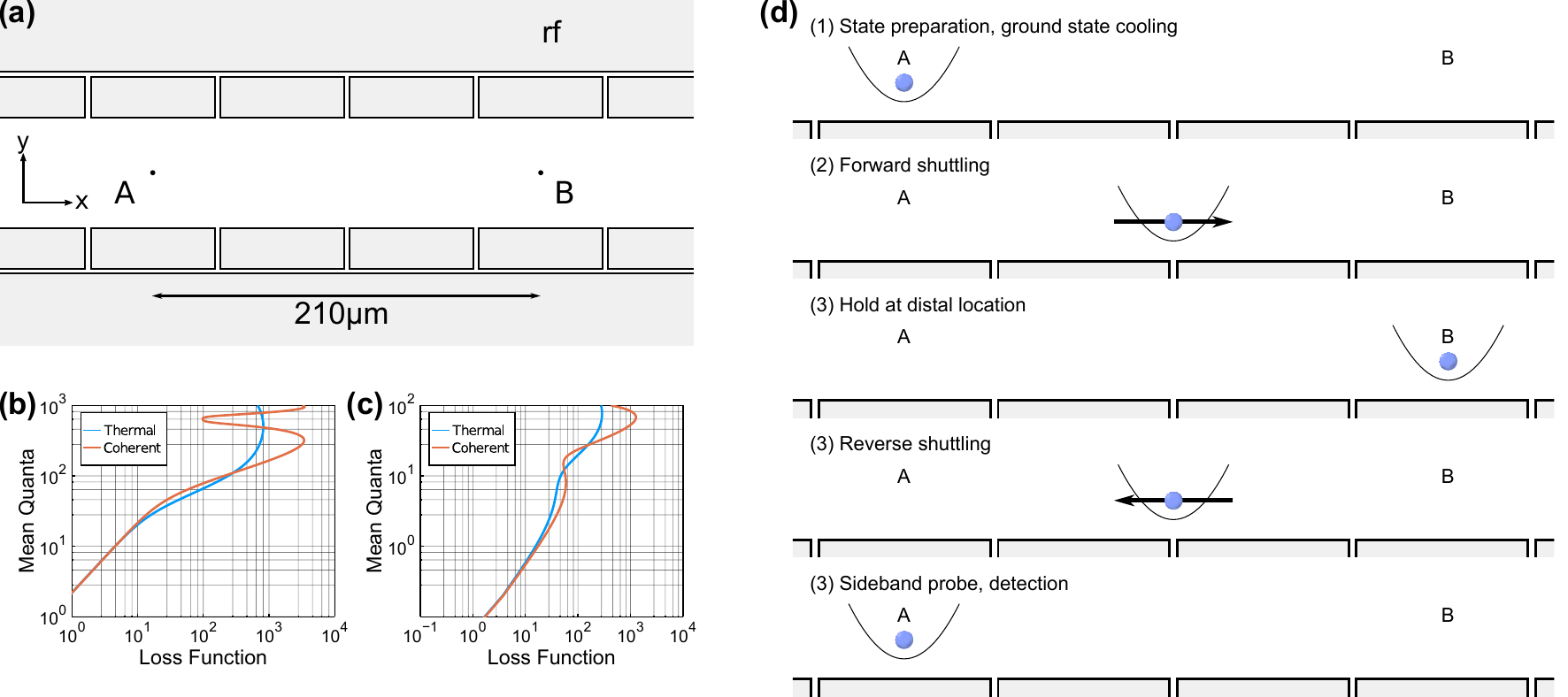}
    \caption{Geometry of the system and loss function.
    (a) An ion is trapped at location $A$ where it is prepared in the axial ($\hat{x}$) ground state of motion and in the electronic state $\ket{S_{1/2}, -\frac{1}{2}}$.
    It is then shuttled to $B$ in 6~$\mu$s and held there for at least 12~$\mu$s before being shuttled back.
    The drawing shows the rf rails and interior control electrodes (but not the outer control rails).
    The relationship of the loss function to the mean quanta for both thermal and coherent motion is shown in (b) and (c) using experimentally relevant probe times.
    Probe time $t_{m}$ corresponds to the $m$-th order red sideband; (b) shows the relationship for $(t_{1}, t_{2})$ = (3~$\mu$s, 10~$\mu$s) and (c) shows the relationship for (25~$\mu$s, 45~$\mu$s).
    (d) A cartoon of the experimental transport sequence. The ion is prepared at $A$, then shuttled to location $B$ and held there for a variable hold time before returning to $A$ for a probe on the sidebands and detection.}
    \label{fig:geometry}
\end{figure*}

In the experiment, the ion is initially trapped at location $A$ (see Fig.~\ref{fig:geometry}a) with an axial frequency of $\omega_x/2\pi = 2.5~\mathrm{MHz}$ and radial frequencies $\omega_{1,2}/2\pi = (5.6, 6.0)~\mathrm{MHz}$. 
At this location, the ion is spin-polarized to the electronic state $\ket{0} = \ket{S_{1/2}, -\frac{1}{2}}$
and the axial motion is sideband cooled to the motional ground state (mean quanta $\bar{n} \approx 0.03$).
Next, using a transport waveform derived from the optimization state, the ion is shuttled three electrodes to the distal location $B$, 210~$\mu$m away, corresponding to a separate and independent trapping site.
The ion is held at $B$ for a variable dwell time of at least 12~$\mu$s before being returned to $A$ with the reversed waveform. 
After the round-trip transport, the amount of axial excitation is probed through a frequency scan over the first- and second-order red sideband on the $\ket{S_{1/2}, -\frac{1}{2}} \leftrightarrow \ket{D_{5/2}, -\frac{5}{2}}$ ($\ket{0}\leftrightarrow\ket{1}$) transition.
The loss function is a combination of the integrals of these sidebands (Eq.~(\ref{eqn:lossfunction})), which acts as a pseudo-energy as it can be related to the mean quanta of excitation up to hundreds of quanta (Figures \ref{fig:geometry}b and ~\ref{fig:geometry}c). 
A cartoon of the procedure is illustrated in Fig~\ref{fig:geometry}d.
While similar pseudo-energies have been used in other shuttling experiments~\cite{waltherControlling2012,kaushalShuttlingbased2020}, the integrals serve as quick and straightforward proxy measurements of the excitation.

\begin{figure}[tbp]
    \centering
    \includegraphics[width=\columnwidth]{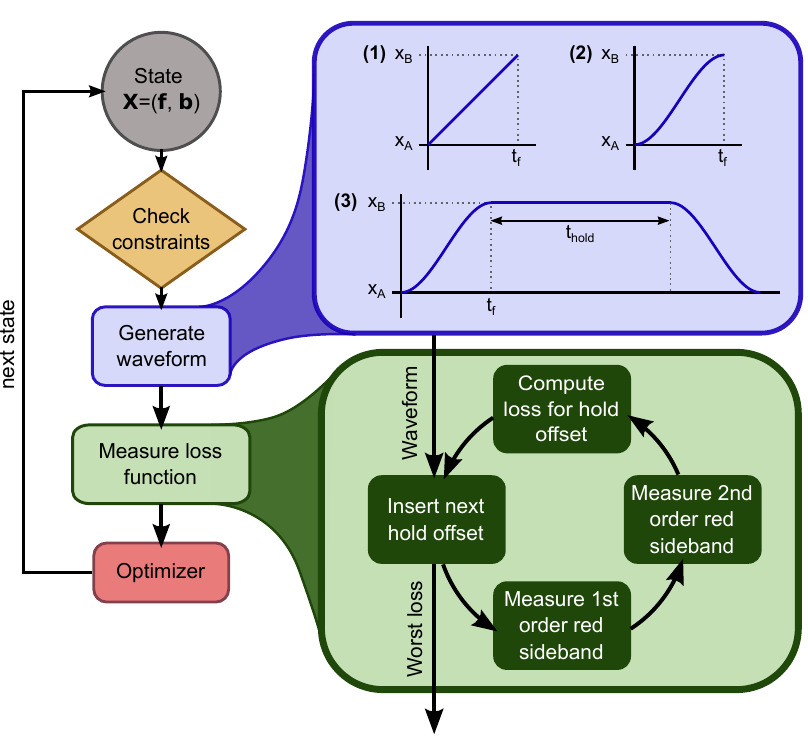}
    \caption{Optimization procedure. 
    Given an optimization state $\mathbf{X}=(\mathbf{f}, \mathbf{b})$, the constraints on the state are first checked and if satisfied are used to generate the waveform from the base solution (1).
    This is achieved by scaling the axial frequency and modifying the trajectory to realize the forward transport solution (2).
    The full waveform consists of the forward solution followed by a hold and then followed by the time reversal of the forward waveform to return the ion to position $A$ (3).
    The loss function consists of a loop over a set of DAC offset steps that are inserted into the hold time.
    The loss is calculated for each offset by measuring the first- and second-order red sidebands.
    The final loss is the worst performing of these offset losses and is sent to the optimizer which generates the next trial state.
    }
    \label{fig:procedure}
\end{figure}

Given a constant axial frequency during transport, the final state of motion is a coherent state of motion~\cite{Reichle2006,lauDecoherence2011}. 
This quantum control problem can be reduced to a classical control problem through unitary transformations~\cite{Reichle2006,hoganLightpulse2009}.
This coherent excitation of the ion at the distal location is revealed by changing the dwell time of the ion at $B$, as it shows a periodicity in the final excitation.
Therefore, these measurements are repeated with the same waveform, with an additional offset to the hold time at the distal location inserted.
These time offsets are chosen to equally sample the phase of axial motion at the distal location.
The value of our loss function for the optimization state, $\mathbf{X}$, is the worst performing of all the hold offsets $h$,
\begin{widetext}
\begin{equation}
    \label{eqn:lossfunction}
    \mathcal{L}(\mathbf{X}) = \max_{h} \mathcal{L}(\mathbf{X} \vert h) = 
	\max_{h} \left[ \alpha_{1} \int_{B_{1}} \!\frac{\mathrm{d}\Delta_{1}}{2\pi} r_{1}(\Delta_{1} \vert \mathbf{X}, h) 
	+ \left( \alpha_{2} \int_{B_{2}}\!\frac{\mathrm{d}\Delta_{2}}{2\pi} r_{2}(\Delta_{2} \vert \mathbf{X}, h) \right)^{2} \right]
\end{equation}
\end{widetext}
where $r_{m}(\Delta_{m} \vert \mathbf{X}, h)$ is the $m$-th order red sideband for hold offset $h$ given the state $\mathbf{X}$.
The hyperparameters $\alpha_{m}$ are chosen to be the same, $\alpha_{1} = \alpha_{2} = 2~\mathrm{kHz}^{-1}$.
For an ion in the state $\ket{0}\!\bra{0} \otimes \rho(\mathbf{X},h)$ after transport, the sideband lineshapes can be shown to be after a probe time $t_{m}$,
\begin{equation}
    \label{eqn:rsb}
	r_{m}(\Delta_{m} \vert \mathbf{X}, h) = \sum_{n=m}^{\infty} \rho_{nn}(\mathbf{X}, h) 
	\frac{4|g_{nm}|^{2}}{\Omega_{nm}^{2}} \sin^{2} \left( \frac{\Omega_{nm} t_{m}}{2} \right),
\end{equation}
where $\Omega_{nm} = \sqrt{4|g_{nm}|^2 + \Delta_{m}^{2}}$ is the detuned Rabi frequency, $\Delta_{m}$ is the detuning from the $m$-th sideband, and the sideband coupling strengths are $g_{n1} = -i\eta g_{0}\sqrt{n}$ and $g_{n2} = - \frac{1}{2}\eta^{2}g_{0}\sqrt{n(n-1)}$ (here, $g_{0}$ is the coupling strength to the carrier transition, and $\eta$ is the Lamb--Dicke parameter).
A quadratic penalty for the integral of the second-order sideband emphasizes the minimization of the second-order sideband before minimizing the first-order sideband.
In figures~\ref{fig:geometry}b and ~\ref{fig:geometry}c, the relation between the value of the loss function and mean quanta are plotted for both the case of a thermal and coherent excitation for the probe times used for a particular experiment.
The resulting measurement of the total loss is passed to the optimizer, which is chosen to use the Nelder--Mead algorithm as it is a simple derivative-free algorithm.
The optimizer then determines the next state $\mathbf{X}$, which is then passed to the experiment for voltage synthesis and test.
Fig.~\ref{fig:procedure} illustrates the optimization procedure.

The initial probe times for the first- and second-order red sidebands were chosen to be slightly less than the effective $\pi$-times so as to not overdrive the sidebands, which would saturate the measurement.
After 150 function evaluations of the Nelder--Mead optimizer, the loss function became insensitive to improvement and leveled off (Fig.~\ref{fig:data}a).
Therefore a second round of the optimization was performed using longer probe times and starting from the final state of the first stage.

\subsection{Optimization state and waveform generation}
For each optimization state $\mathbf{X}$ in the experiment, the shuttling waveform is synthesized just prior to test. 
The waveforms are all derived from a base trapping voltage set which consists of $211$ individual trapping solutions equally spaced along the $210~\mu$m path from $A$ to $B$. 
Each solution is generated through a constrained optimization problem to generate the least-norm voltage array with a fixed $2.5~\text{MHz}$ axial frequency for a calcium ion with a unique well location along the path.
These solutions are derived with respect to a boundary element model of our device, which yields trapping solutions with axial frequencies within $10\%$ of the experimentally measured value.
First, a forward transport waveform is constructed to transport the ion to the distal location at a speed of $35~\text{m/s}$.
Then the full waveform that is run in the experiment is the concatenation of this forward solution followed by a hold at the distal location for 12~$\mu$s (plus an additional offset) and finally the time reversal of the forward waveform to bring the ion back for measurement. 

The optimization state consists of a list of $n_{f}$ axial frequency points $\mathbf{f} = \{f_{j}\}_{j\in1:n_{f}}$ and $n_{t}$ trajectory control points $\mathbf{b} = \{b_{j}\}_{j\in1:n_{t}}$ to control the axial frequency of the ion along the path and the harmonic well trajectory.
Each frequency point $f_{j}$ fixes the axial frequency at a spatial position $x_{j} = x_{A} + j\delta x$, where $\delta x = (x_{B}-x_{A})/(n_{f} + 1)$ so that they are equally spaced between $A$ and $B$. 
Between these points, the axial frequency is linearly interpolated and each trapping solution in the base solution is scaled by a factor to match the desired frequency at that position.
We constrain the axial frequencies with an exponential penalty for values outside the range $[1.5, 3.5]~\mathrm{MHz}$; this ensures reasonable voltages and a physical trap throughout the shuttling procedure.
The trajectory $s:[0,1] \rightarrow [0,1]$ determines the harmonic well location via $x_{\mathrm{well}}(t) = x_{A} + (x_{B} - x_{A}) s(t/t_{f})$ where $x_{A,B}$ are the spatial location of $A$ and $B$ and $t_{f} = |x_{B}-x_{A}|/v = 6~\mu s$ is the transport time for the desired velocity $v=35~\text{m/s}$.
It is constrained to be symmetric (\emph{i.e.} $s(1-\tau) = 1 - s(\tau)$, $\tau = t/t_f$) and have fixed endpoints with zero initial and final velocities.
The trajectory is defined through a B\'ezier curve, or Bernstein interpolation,
\begin{equation}
    s(\tau) = \sum_{j=0}^{N} s_{j} \binom{N}{j} \tau^{j} (1-\tau)^{N-j} .
\end{equation}
The choice of such an interpolation makes it easy to automatically satisfy the trajectory constraints: $s_{0} = s_{1} = 0$, $s_{N} = s_{N-1} = 1$, and $s_{j} + s_{N-j}=1$.
The trajectory control points correspond to the lowest non-zero B\'ezier coefficients, $b_{j} = s_{j+1}$, resulting in an $N = 2n_{t} + 3$ order polynomial.
The forward waveform is formed by determining the position of the trajectory at each DAC step and determining the voltages from the base solution through linear interpolation.
If the voltage waveform does not exceed the voltage budget of the electronics, the waveform is applied and the lineshapes of the first- and second-order red sidebands are measured.
Violation of the voltage budget also leads to exponential penalty in the value of the loss that is sent to the optimizer.

\subsection{Data}\label{sec:results}
\begin{figure*}[htbp]
    \centering
    \includegraphics[width=\textwidth]{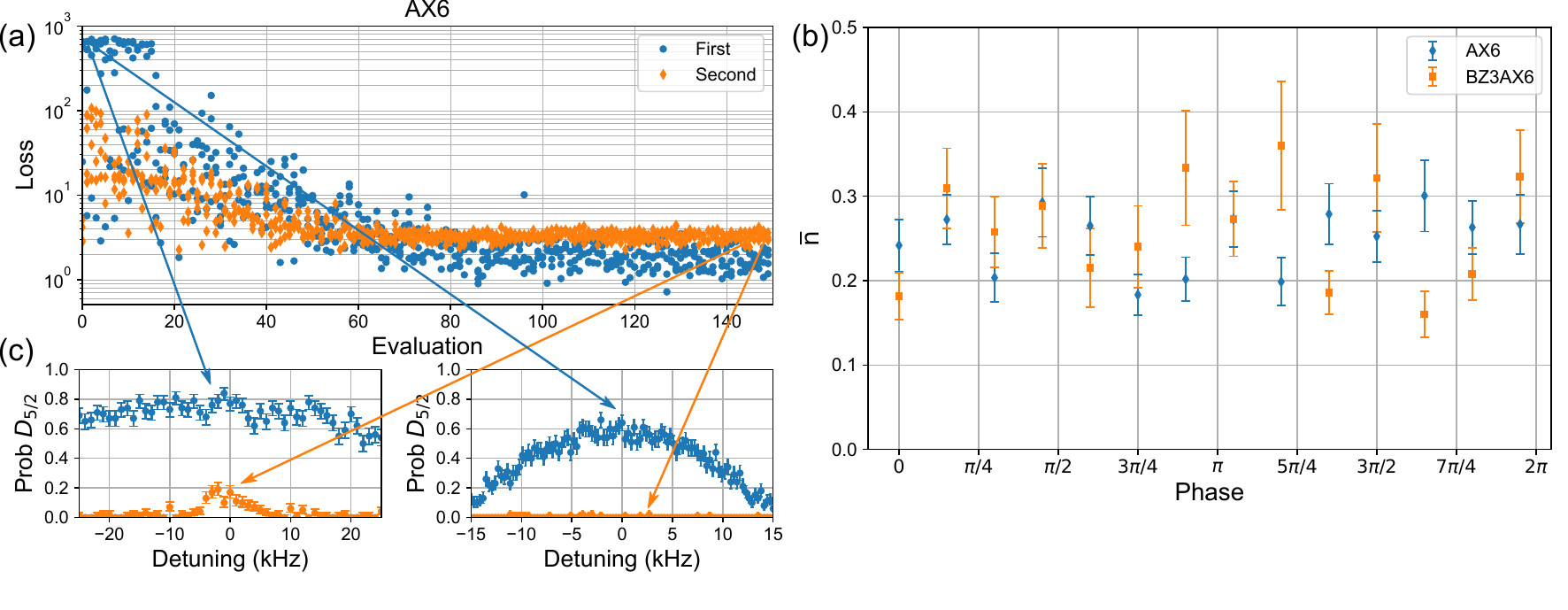}
    \caption{Experimental results of the optimization procedure.
    (a) This plot illustrate how the loss function proceeds over function evaluations for the axial frequency optimization for the first and second stages (blue and orange, respectively). 
    The trajectory and trajectory-plus-axial optimizations display similar behavior.
    For each evaluation of the total loss for a given state $\mathbf{X}$, we plot the loss for each hold offset, the maximum of which is sent to the optimizer.
    (b) Measurement of the mean quanta via sideband thermometry at the end of the optimization for both the axial and trajectory-plus-axial optimization as a function of the phase of the axial motion at the distal location $B$. 
    (c) Representative plots of the first (left) and second (right) order red sidebands for the initial (blue) and final (orange) optimization states. \emph{N.b.} The scans for the final optimization were performed with much longer probe times than the initial state.}
    \label{fig:data}
\end{figure*}

We applied our optimization routine to transport waveforms parameterized by 1) B\'ezier trajectories defined by three control points, 2) linear trajectories defined by six intermediate axial frequencies, and 3) a combination of the two. 
The number of trajectory control points was chosen to provide flexibility while maintaining a reasonable order polynomial (here, $n_{t}=3$ corresponds to a $9$-order polynomial). 
Likewise, the number of axial control points was found to be sufficient for our experimental velocity; other velocities might require a different number of axial control points.
As seen in Fig.~\ref{fig:data}a, a strong periodicity in the integrated sidebands over the various dwell times was observed at the beginning of the optimization, which is to be expected.
However, as the optimization proceeded the performance tended to become uniform across the hold offsets, indicating insensitivity to the dwell time. 
Due to the different probe times, the overall value of the loss functions between different optimizations and runs are not directly comparable without relating the loss to the mean quanta.
The initial $\pi$-time for the optimizations with the B\'ezier parameters were found to be longer than a linear trajectory (used in the axial-control-only optimization), indicating an initial lower motional excitation.

The absolute performance of the resulting optimized waveform was measured using sideband thermometry \cite{diedrichLaser1989}.
Fig.~\ref{fig:data}b shows the quanta gain versus phase at the distal location for the axial control and the trajectory-plus-axial optimization routines, which both exhibit sub-quanta performance.
The final waveform of the trajectory-only optimization did not yield sub-quanta performance and could not be probed reliably with sideband thermometry (a minimum of $1.4\pm0.4$ quanta was observed).
This could be due to an insufficient exploration of the parameter space, inability to escape a local minimum, or an insufficient number of parameters.
A background heating rate of $295 \pm 24$~quanta/s adds a negligible amount of heating (approximately $0.01$ quanta) to the ion over the course of the transport.

\begin{figure}
    \centering
    \includegraphics[width=\columnwidth]{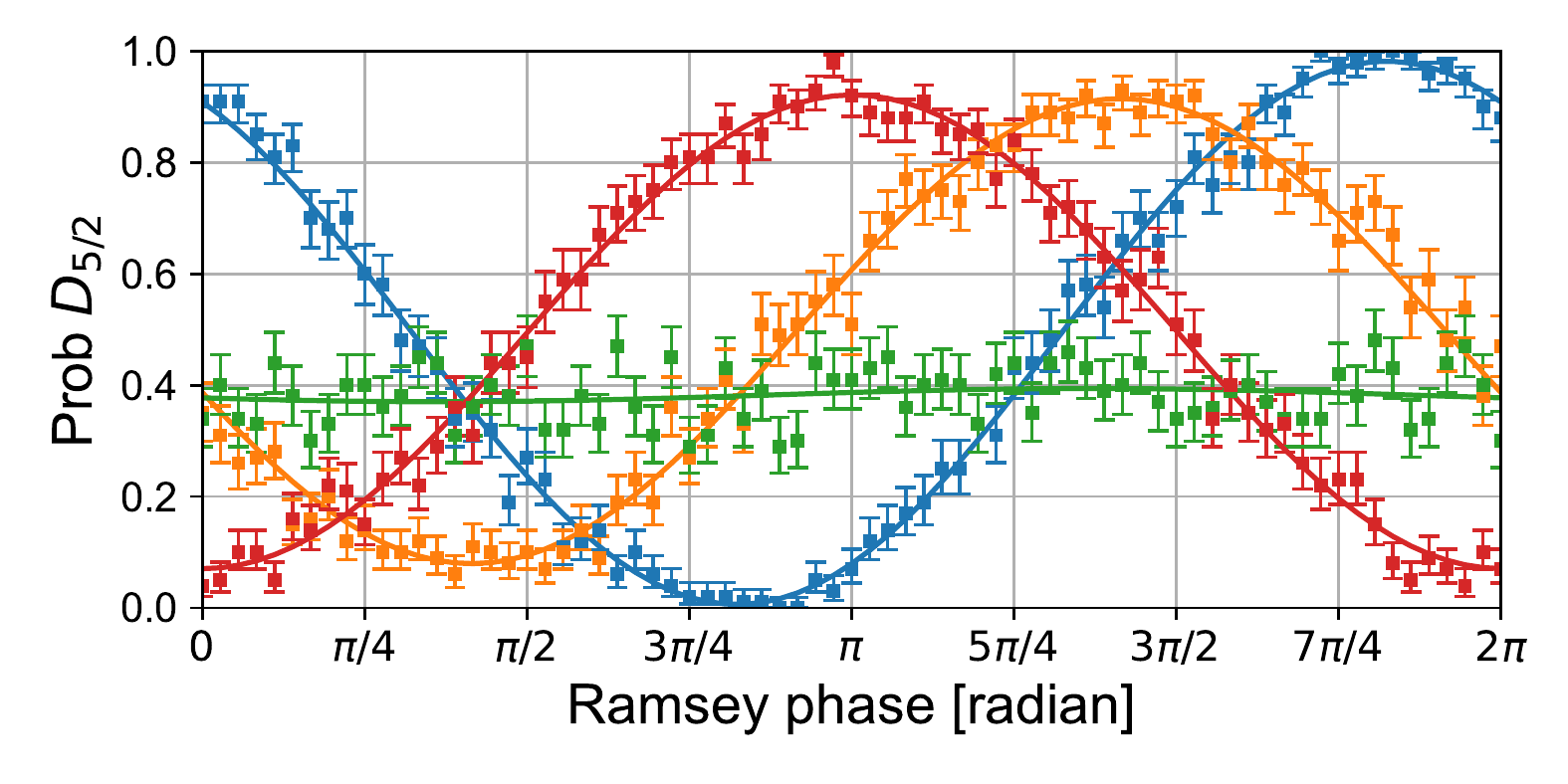}
    \caption{Ramsey phase scans to verify ion transport.
    We illustrate a Ramsey measurement of the ion coherence for four cases.
    The baseline case, with neither shuttling nor probe, is in blue.
    In orange, the ion is shuttled without a probe at $B$, showing that coherence is maintained throughout the procedure.
    To show the probe has no effect when the ion is at $A$, the ion is held stationary while the probe is turned on (red).
    Coherence is only lost (green) when the ion is shuttled to $B$ and the probe is turned on.
    }
    \label{fig:ramsey}
\end{figure}

Since it is possible that the optimization could generate a non-shuttling waveform to achieve low excitation, a Ramsey measurement is used to verify that the ion is transported all the way to the distal location with the optimized waveform (Fig.~\ref{fig:ramsey}).
Prior to shuttling a $\pi/2$ pulse is applied to the ion on the $\ket{S_{1/2},-\frac{1}{2}} \leftrightarrow \ket{D_{5/2}, -\frac{3}{2}}$ quadrupole transition.
Our typical transport and hold shuttling procedure is performed, followed by a final $\pi/2$ pulse.
For verification, we illuminate the distal location with a 397 nm laser resonant with the $S_{1/2}\leftrightarrow P_{1/2}$ dipole transition as a probe to destroy the coherence if the ion is successfully shuttled to that position.
A Ramsey phase scan was performed for the four combinations of shuttling on/off and probe on/off.
We see that only when the probe is on and the ion is transported that coherence is lost.

\begin{figure}
    \centering
    \includegraphics[width=\columnwidth]{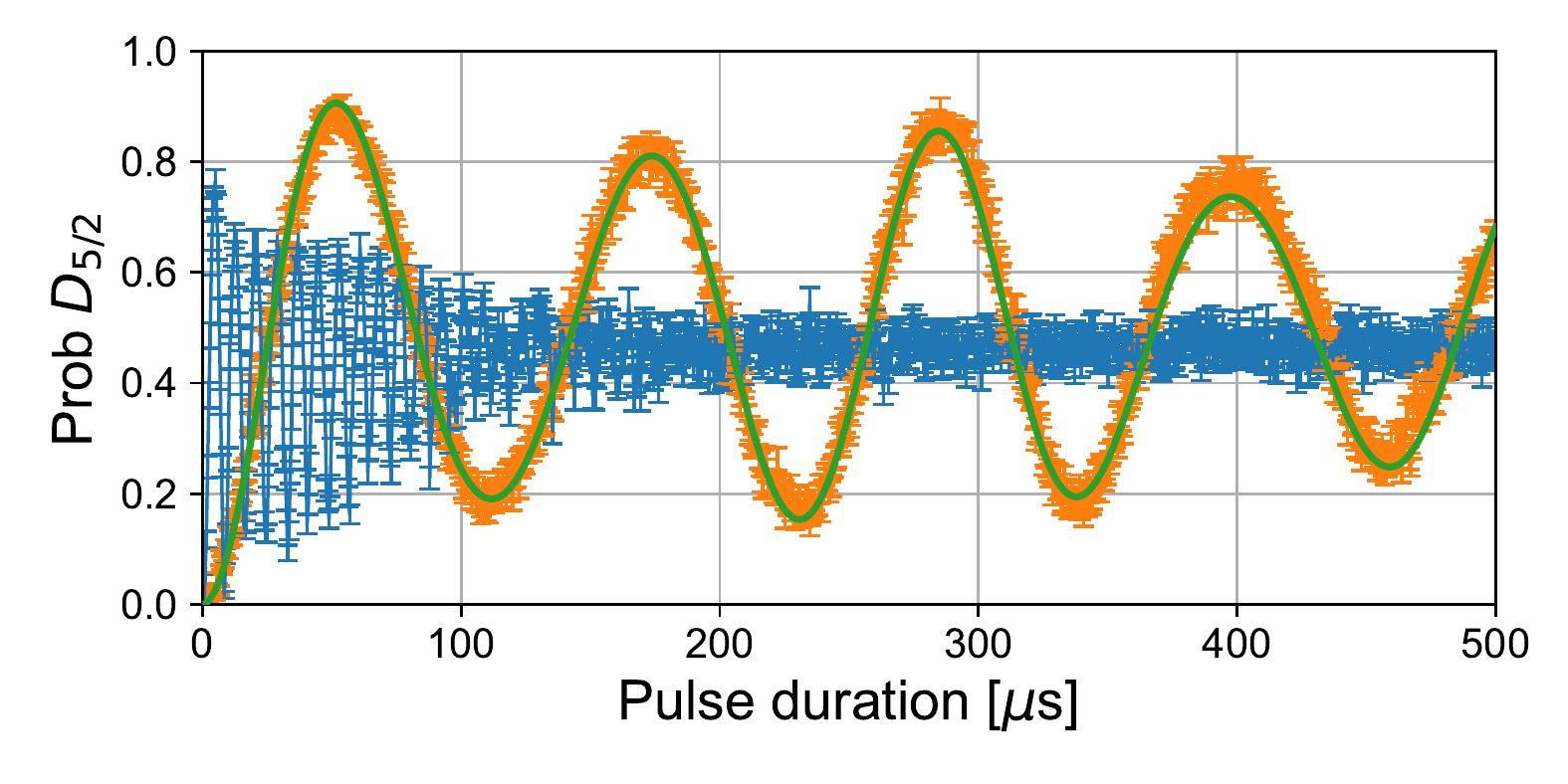}
    \caption{Blue sideband Rabi oscillations after transport for the optimized decimated (blue) and non-decimated (orange) trajectory-plus-axial optimizations.
    The green curve is a maximum likelihood estimation for a displaced thermal state illustrating sub-quanta excitation.
    For the decimated version the ion is in a highly excited state after transport and did not achieve sub-quanta transport.
    }
    \label{fig:decimated}
\end{figure}

To determine whether the high speed electronics are necessary at this transport velocity, a slower-speed DAC was emulated by decimating the trial waveform and upsampling it through a zero-order hold. 
The same trajectory-plus-axial optimization procedure as above was applied, achieving only minimal improvement over the initial transport. 
No waveform was generated with the same number of function evaluations that could achieve sub-quanta excitation with these artificially slow electronics.
The resulting waveform from the decimated version left the ion highly excited; a long blue-sideband Rabi measurement was performed and found consistent with a highly excited motional state.
Fig.~\ref{fig:decimated} shows this data in comparison to a similar measurement for the trajectory-plus-axial optimized waveform.
Fitting the trajectory-plus-axial Rabi oscillation data to a displaced thermal state results in a mean quanta of $\bar{n}_{\mathrm{coh}} + \bar{n}_{\mathrm{th}} = 0.26$, in agreement with the sideband thermometry measurement ($0.31 \pm 0.08$ quanta for this particular hold offset).
It was difficult at this sub-quanta level to discern the relative contributions of the thermal and coherent excitation from a maximum likelihood estimate fit to the Rabi data, as this method is only sensitive to motional state populations and not coherences between the motional states.

\section{Discussion}
Another approach that was considered for optimizing the waveform involved pre-compensating the voltage waveform to account for the low-pass filter attached to the chamber~\cite{bowlerArbitrary2013}.
This approach was investigated in the context of the optimization procedure through the insertion of a digital filter prior to analog voltage synthesis. 
In this approach, the filter coefficients served as the optimization state.
We found an infinite impulse response (IIR) filter to generally be unstable, while an inherently stable FIR filter consistently resulted in voltage waveforms exceeding the range of the DACs during an attempted optimization.
This approach was therefore unsuitable for the optimization loop since most of the optimization was spent in a region that violated the voltage budget.
Although a more complex constrained optimization procedure might be suitable for the pre-compensation approach, the performance of the trajectory and frequency control indicates that such complexity is unnecessary to acheive sub-quanta shuttling.

This closed-loop optimization procedure was used to generate voltage waveforms that transport an ion across multiple electrodes to a separate trapping well at a speed of 0.5 electrodes per microsecond, or $35~\text{m/s}$.
By tuning three trajectory parameters that define a B\'ezier curve and six axial frequencies at discrete points in the ion's path, the motional excitation following transport was limited to $0.36\pm0.08$ quanta.
The B\'ezier potential well trajectory initially performed better than a linear trajectory, however we suspect that it was close enough to a local minimum such that the local optimizer was unable to find a sub-quanta solution.
On the other hand, we observed that axial frequency control alone was sufficient for sub-quanta transport. 
At a lower sample rate, the optimization procedure was unable to make sufficient progress in developing a suitable voltage waveform, indicating that high-speed electronics are an enabling technology for fast shuttling.
This hardware and optimization technique can potentially be applied in other transport situations (split/join or junction shuttling) as well as in larger ion traps with greater site-to-site variation or even other quantum technologies, where human-in-the-loop tuning is impractical.

\section{Methods}
\subsection{Experiment}
A single calcium ion is trapped in the linear section of a High-Optical-Access surface-electrode radiofrequency (rf) Paul trap \cite{blainHybrid2021}.
The ion is trapped approximately 70~$\mu$m above the surface of the trap and is tightly confined in the transverse direction by applying a 140~V amplitude rf signal at 51~MHz to rf rail electrodes.
Axial confinement is provided through voltages applied to specific interior control electrodes which have a pitch of 70~$\mu$m while a 60~$\mu$m gap in the substrate provides an open slot below the ion (see Fig.~\ref{fig:geometry}a).
Control rails outside the rf electrodes provide principal axis rotation in the transverse plane. 

Permanent magnets provide a 9.5~G vertical magnetic field in order to split the ground-state Zeeman levels $\ket{S_{1/2}, \pm\frac{1}{2}}$.
Doppler cooling and state readout are performed with a laser addressing the $S_{1/2}\leftrightarrow P_{1/2}$ transition at 397~nm, while internal state manipulation of the ion is performed with a narrow 729~nm laser addressing the $S_{1/2}\leftrightarrow D_{5/2}$ transition.
Additionally, two repump lasers at 866~nm ($D_{3/2}\leftrightarrow P_{1/2}$) and 854~nm ($D_{5/2}\leftrightarrow P_{3/2}$) are used. 
The ion is spin-polarized to the $\ket{0} = \ket{S_{1/2}, -\frac{1}{2}}$ state by coherently driving the $\ket{S_{1/2}, +\frac{1}{2}} \leftrightarrow \ket{D_{5/2}, m'}$ transition, followed by a repump pulse to the $P_{3/2}$ manifold and spontaneous emission back to the $S_{1/2}$ manifold.
Ground state cooling of axial motion is achieved in a similar fashion, where the coherent drive is tuned to the motional red-sideband of the $\ket{S_{1/2}, -\frac{1}{2}}\leftrightarrow \ket{D_{5/2}, -\frac{5}{2}}$ transition.

After shuttling, the first- and second-order red sidebands are probed by pulsing the 729~nm laser for a given probe time $t_{1}$ and $t_{2}$ prior to applying the detection beam at 397~nm. 
The lineshape of the $m$-th order red sideband, $r_{m}(\Delta, \mathbf{X})$ (Eq.~(\ref{eqn:rsb}), is defined as the probability of the ion transitioning to the state $\ket{1} = \ket{D_{5/2}, -\frac{5}{2}}$,
\[
r_{m}(\Delta, \mathbf{X}) = \tr\left[ \ket{1}\!\bra{1} \chi_{m}(\Delta_{m}, \mathbf{X})\right]
\]
where $\chi_{m}(\Delta, \mathbf{X})$ is the density matrix of the qubit--motion coupled system after transporting and then probing the $m$-th order red sideband with a laser detuned by $\Delta_{m}$ from the sideband for a duration $t_{m}$. 

\subsection{Lineshape}
To calculate the lineshape, we assume the ion is initially in the state $\chi(0) = \ket{0}\!\bra{0} \otimes \rho(\mathbf{X})$ immediately after transport and prior to the probe.
The system coherently evolves under the probe according to the Hamiltonian
\begin{equation}
    H = \omega_{t} a^{\dagger}a - \Delta_{0} \sigma^{\dagger}\sigma + g_{m}\sigma a^{m\dagger} + g_{m}^{*}\sigma^{\dagger}a^{m}
\end{equation}
where the sidebands are well-resolved and the ion is in the Lamb--Dicke regime.
Here, $a$ corresponds to the phonon annihilation operator for the axial mode and $\sigma =
\ket{0}\!\bra{1}$.
The detuning $\Delta_{0} = \omega_{L} - \omega_{0} = \Delta_{m} - m \omega_{t}$ is the detuning of the laser from the carrier
transition, which is expressed in the second equality in terms of the detuning from the $m$-th order
red sideband, $\Delta_{m}$. 
The sideband coupling strengths are given by
$g_{m} = \frac{(-i\eta)^{m}}{m!} g_{0}$, where $g_{0}$ is the carrier coupling strength.

The excitation operator $\hat{N} = a^{\dagger} a + m \sigma^{\dagger}\sigma$ is a conserved quantity for this Hamiltonian, and thus any eigenstate of $\hat{N}$ preserves the excitation number. 
This results in a collection of closed manifolds whose dynamics are independent of one another. 
The dimensionality of each manifold is either one-dimensional (for eigenvalues $n < m$) or two-dimensional (eigenvalues $n \ge m$).
For the case $n<m$, the basis state is $\ket{0, n}$, while for $n\ge m$ the states $\ket{0,n}$ and $\ket{1, n-m}$ are coupled. 

Each subspace can be diagonalized (for $n\ge m$) in terms of the dressed states, 
\begin{align*}
    \ket{n,+}  &= \cos \frac{\theta_{n}}{2} \ket{1, n-m} + e^{i \arg g_{m}} \sin \frac{\theta_{n}}{2} \ket{0, n} \\
    \ket{n,-}  &= -e^{-i \arg g_{m}} \sin \frac{\theta_{n}}{2} \ket{1, n-m} + \cos \frac{\theta_{n}}{2} \ket{0, n} 
\end{align*}
where $\tan \theta_{n} = -2|g_{nm}|/\Delta_{m}$.
Here we define the $m$-th order sideband coupling strength for the $n$-th manifold, $g_{nm} = g_{m}
\sqrt{n!/(n-m)!}$.
These states are eigenstates of the Hamiltonian with eigenenergy 
\[
    \epsilon_{\pm} = \frac{-1}{2} \left[ \Delta_{m} \pm \sqrt{ \Delta_{m}^{2} + 4|g_{nm}|^{2} } \right]
\]
The probability for measuring the ion in $\ket{1}$ after probing the $m$-th sideband is found by expressing the initial state in terms of the dressed states, applying the phase accumulated over time $t_{m}$, projecting onto $\ket{1}$, and taking the trace. Such a
procedure yields Eq.~(\ref{eqn:rsb}),
\begin{equation*}
    r_{m}(\Delta_{m}) = \sum_{n=m}^{\infty} \rho_{nn}
\sin^{2}\theta_{n} \sin^{2} \left( \frac{t_{m}}{2} \sqrt{ \Delta_{m}^{2} + 4|g_{nm}|^{2} } \right)
\end{equation*}
where  $\sin^{2}\theta_{n} = \frac{4|g_{nm}|^{2}}{ \Delta_{m}^{2} + 4|g_{nm}|^{2} } $, 

\subsection{Electronics}
The voltages applied to the trap electrodes are derived from a custom 96 channel high-speed arbitrary waveform generator.
Signals are specified in 30~ns steps and filtered on the digital side with finite impulse response filters that result in a normalized pass band frequency of $12~\mathrm{MHz}$ and a stop band frequency of $15~\mathrm{MHz}$ with greater than $100~\mathrm{dB}$ attenuation.   

A DAC for each channel outputs a $\pm 2.5~\mathrm{V}$ signal which is amplified to $\pm 10~\mathrm{V}$ by a power amplifier with low-distortion and high-speed current-feedback. Anti-alias low-pass filters are used to reject unintended signal generation in higher-order Nyquist domains and yield a $12~\mathrm{MHz}$ analog bandwidth.  These voltages are then delivered through sixth-order low-pass filters at the vacuum chamber feedthrough with $3~\mathrm{dB}$ cutoff at $1.3~\mathrm{MHz}$, in order to reduce heating from electrical noise at the axial frequency. The timing of the system is governed by a temperature-compensated voltage-controlled crystal oscillator that is phase-locked to an external 10 MHz reference clock.

\section{Data Availability}
The data presented in this manuscript are available from the corresponding author upon reasonable request.

\section{Author contributions}
J.~D.~S., H.~M., L.~P.~P. and D.~S. conceived and carried out the experiment. 
H.~C., J.~G., V.~H., J.~L., D.~M. developed the high speed electronics and control software. 
J.~V.~D.~W. provided the base voltage solutions and experiemntal control software.

\section{Competing Interests}
The authors declare no competing interests.

\section{Acknowledgements}
This research was funded by the U.S. Department of Energy, Office of Science, Office of Advanced Scientific Computing Research.
Sandia National Laboratories is a multimission laboratory managed and operated by National Technology \& Engineering Solutions of Sandia, LLC, a wholly owned subsidiary of Honeywell International Inc., for the U.S. Department of Energy’s National Nuclear Security Administration under contract DE-NA0003525.  
This paper describes objective technical results and analysis. 
Any subjective views or opinions that might be expressed in the paper do not necessarily represent the views of the U.S. Department of Energy or the United States Government.


\end{document}